\documentclass[10pt,a4paper]{article}
\usepackage[T1]{fontenc}
\usepackage{inputenc}
\usepackage[english]{babel}
\usepackage[nohead,includefoot,margin=2cm]{geometry}
\usepackage[compact,raggedright,small]{titlesec}
\usepackage[affil-it]{authblk}
\usepackage[runin]{abstract}
\usepackage{multicol}
\usepackage{blindtext}

\usepackage{graphicx} %eps figures can be used instead
\usepackage{epsfig}
\usepackage{epstopdf}
\usepackage{setspace}
\usepackage{oldgerm}
\usepackage{float}
\usepackage{amsmath}
\usepackage{amssymb}

\graphicspath{{images/}}

\newcommand{\tauLB}{{\tau_{\rm{LB}}}}
\newcommand{\kB}{{k_{\rm B}}}

\addto{\Affilfont}{\small}

\title{\large\bfseries Polyelectrolytes polarization in non-uniform electric fields}
\date{\small \date{}}
\author[1]{Farnoush Farahpour}
\author[2]{Fathollah Varnik}
\author[1]{Mohammad Reza Ejtehadi\footnote{e-mail: ejtehadi@sharif.edu}}
\affil[1]{Sharif University of Technology, Department of Physics, P.O. Box 11155-9161, Tehran, Iran. Fax: +98 21 66022711; Tel: +98 21 66164501.}
\affil[2]{ICAMS, Ruhr-Universität Bochum, IC 02-505, Universitätsstr. 150, 44801 Bochum, Germany.}

\begin{document}
\maketitle

\begin{abstract}
Stretching dynamics of polymers in microfluidics is of particular interest for polymer scientists. 
As a charged polymer, a polyelectrolyte can be deformed from its coiled equilibrium configuration to an extended chain by applying uniform or non-uniform electric fields. By means of hybrid lattice Boltzmann-molecular dynamics simulations, we investigate how the condensed counterions around the polyelectrolyte contribute to the polymer stretching in inhomogeneous fields. As an application, we discuss the translocation phenomena and entropic traps, when the driving force is an applied external electric field.
\end{abstract}

\vspace{1cm}
\section{Introduction}

\newcommand{\myepsilon}{\text{\usefont{OML}{cmr}{m}{n}\symbol{15}}}

Electrohydrodynamical behavior of polyelectrolytes (PEs) in electric fields has drawn scientists' attention in the last decades. Migration of charged polymer chains under the action of electric field, occurs in many \textit{in vivo} and \textit{in vitro} phenomena and experiments such as RNA migration from nucleus' pores, electrophoresis techniques and translocation experiments\cite{1}. PEs are deformable charged macromolecules and these two attributes are sufficient to make their mechanical behavior to be highly complex and investigating the aspects of their dynamics to be very challenging.

Mechanical response and deformation of polymer chains in different situations have been vastly studied in literature\cite{2}. Constraints, flows and external forces can change the configuration of a polymer from its natural coiled shape to an extended or compressed form\cite{2,3,4,5}. When a polymer carries charge on its monomers, an electric external field can be used to deform it\cite{6,7}.

Due to their electrostatic characteristics, PEs exhibit distinct electrophoretic behaviors, which are, however, poorly understood. While counterion (CI) condensation and effective charge of the chains are still subjects of investigations, yet we know that condensed CIs have a complicated role, which goes beyond a simple screening of the charge of the chain \cite{8}.

One of the simplest situations for interaction of a PE and electric fields is its free migration in a homogeneous electric field, called capillary electrophoresis\cite{9}. It has been shown that electroosmotic flow of condensed CIs in Debye length around PE's backbone eliminates the hydrodynamic interactions (HI) in moderate electric fields, that raises to electrophoretic velocities which are independent of PE's length\cite{10,11}. 

Efforts have been done to overcome this deficiency of capillary electrophoresis by using microchip electrophoresis or designing microfluidic devices such as entropic traps. In a big group of such new methods, PEs are subjected to inhomogeneous electric fields (IEF), at least in some part of the device\cite{2,12}.

When PEs are moving in an IEF, they can be stretched in the direction of electric field, exactly in the same way as their migration in extensional flows\cite{2,7}. Many groups have tried to describe the behavior of PEs in such fields, using electrohydrodynamic equivalence. It has been shown that having electric field gradient tensor, one can predict the extension axis of PEs in non-uniform electric fields\cite{7,13}.

One important issue here is the role of this deformation on the length-dependent mobility of PEs. Progress along this line is of great interest for nanofluidic systems as a candidate for gel electrophoresis substitution. In a recent study, it was shown that both the double-layer polarization and the CI condensation can be influenced significantly by the shape of an ellipsoidal PE. This makes the mobility of the chain depend on its charge and shape, in aqueous solutions\cite{14}.

On the other hand, mobile but constrained ions in a sufficiently large electric field or an IEF can be rearranged and make the whole PE complex, polarized. Simulation studies show that this effect can yield to elongation of PEs and so change their dynamical behaviors \cite{15}. A theoretical and numerical study showed that, in addition to the electroosmotic flow around a PE, polarization effects yield profound and interesting electrophoretic behaviors of PEs \cite{16}.
 
Studying the dynamics of polymers in aqueous solution has another difficulty which is arisen due to the long-range HI between the monomers of the chain which gives rise to complicated behaviors that can not be neglected in a wide range of experiments\cite{11,13,17}.

The complicated nature of PEs' dynamic with all the above mentioned characteristics makes it practically impossible to consider all the features of their dynamics in a theoretical study so that simplifications are always necessary. In this context, computer simulations play an important role as they provide detailed information on the microscopic dynamics of PEs\cite{7,10,13}.

Impact of HI, polarization effects and inhomogeneity of electric field on dynamics of flexible PEs in microfluidic systems have been studied in several articles but to the best of our knowledge details of the phenomenon and distinct relevant parameters have not been addressed so far. In a previous study, we examined the dynamical behavior of a flexible ssDNA in IEF of a nanopore and showed that polarization effects and electroosmotic flow of CIs must be acccounted for, when studying the dynamical behavior of PEs.\cite{18}.

In this study, we employ a hybrid simulation approach which combines a molecular dynamics (MD) integrator and the Lattice Boltzmann (LB) method to investigate the effect of CI rearrangement and polarization in gradients of electric fields. Hydrodynamic interactions, which can play a role in out-of-equilibrium dynamics of polymers, are explicitly accounted for by the LB solver. Two different systems with different designs are considered to establish IEFs as the driving force for the migration of the chain.

\vspace{1cm}
\section{Model and Method}

We employ MD-LB hybrid simulations using the ESPResSo package \cite{19} to study the behavior of linear PEs of different lengths in IEFs. To take into account both the hydrodynamic and electrostatic interactions, we use a hybrid mesoscale method in which the LB method \cite{20,21,22} is applied to simulate the HI effects in the system and a P3M algorithm is used to calculate the full electrostatic interactions between charged monomers and CIs \cite{23}. To investigate the polarization effect, systems with and without explicit ions are simulated. When there is no CI or coion and so no charge balance in the system, we use the Debye-H\"uckel potential.

A bead-spring coarse-grained approach is used for the MD part of the simulation. The non-bonded interactions between beads as well as between the beads and the walls are modelled by the Weeks-Chandler-Anderson (WCA) version of the Lennard-Jones potential with the standard energy depth, $\myepsilon$, and length, $\sigma$, parameters which are the length and energy units of the system, respectively. Unless explicitly stated, all the quantities in this paper are given in these reduced units. If we consider single stranded DNA (ssDNA) which is a flexible PE as a case study, an approximate map of the present model to a physical system leads to the real units of $\myepsilon=0.59~\rm{Kcal/mol}$, $\sigma=1.0~\rm{nm}$.

The bond between the adjacent beads in the chain is modelled via a finitely extensible nonlinear elastic (FENE) potensial: $U_{\text{FENE}}(r) =-0.5 k R^{2}\ln[1-(r/R)^2]$\cite{18} with $k=30$ and $R=1.5$. Ions (CIs and coions) have a radius of $d_{\rm ion}=0.425$ \cite{18} and
all simulations are performed at a temperature of $\kB T=1$.

In all the studied cases with explicit ions, an appropriate amount of monovalent CIs is added to the charged PE to ensure charge neutrality. In order to reach a desired salt concentration, a corresponding number of coions and CIs is added to the system. The salt concentration in this study is equal to $75mM$ and is selected so that the artifacts of the periodicity are negligible \cite{18}.
Each monomer carries a reduced electric charge of $-3$ and each CI and salt ion has a charge of $\pm1$. 
The Bjerrum length is $l_{\rm B}=e^{2}/4 \pi \myepsilon_0 \myepsilon_{\rm w} k_{\rm B}T=0.70$.

Simulations are performed in a periodic rectangular box at a constant monomer concentration of approximately $5mM$. The system size and geometry are adapted depending on the situation investigated.

In all of the simulations reported here, HI are included via a frictional coupling of the particle dynamics to the fluctuating LB method. We use the D3Q19 model (with 19 discrete velocities) implemented in the package. At each time step, the LB particles undergo local collision at the lattice sites, and the evolution of $n_i$, is governed by the LB equation\cite{20,21}:
\begin{equation}\label{LB_update}
n_i(\textbf{r}+\textbf{c}_i\Delta \tau,t+\Delta \tau) = n_i(\textbf{r},t)+\sum_{j=1}^b L_{ij}[n_j(\textbf{r},t)-n_j^{eq}(\rho,\textbf{u})]+n'_i(\textbf{r},t)
\end{equation}
where $\textbf{r}$ is the lattice position, $\textbf{c}_i$ is the discrete velocity, $t$ is the fluid propagation time, $\Delta \tau$ is the LB time step, $b$ depends on the details of the model (here is equal to 19). L is a collision operator for dissipation due to fluid particle collisions such that the fluid always relaxes toward the local pseudo-equilibrium distribution $n_j^{eq}(\rho,\textbf{u})$ that depends on the local hydrodynamic variables, fluid density, $\rho=\sum_i n_i$, and velocity, $\textbf{u}=\sum_i n_i\textbf{c}_i/\rho$. The last term, $n'_i(\textbf{r},t)$, is purely stochastic and is essential in simulating thermal fluctuations that drive Brownian motion. In this study, thermal fluctuations are included in the non-hydrodynamic modes as well as the hydrodynamic modes in order to satisfy the fluctuation-dissipation theorem not only on hydrodynamic but on all length scales\cite{20}.

At equilibrium, the velocity distribution functions can be represented as a second-order expansion of the Maxwell-Boltzmann distribution, given by
\begin{equation}\label{equilibrium_distribution}
n_i^{eq}(\rho,\textbf{u})=\rho w_{c_i}(1+\frac{\textbf{c}_i\cdot\textbf{u}}{c_s^2}+\frac{(\textbf{c}_i\cdot\textbf{u})^2}{2c_s^4}-\frac{u^2}{2c_s^2}),
\end{equation}
where $w_{c_i}$ are weight factors depending on the lattice and $c_s$ is the speed of sound.

MD and LB are coupled through the friction force  $\textbf{F}_{\text{friction}}=-\Gamma_{\text{bare}}(\textbf{u}_{\text{monomer}}-\textbf{u}_{\text{fluid}})$ where $\Gamma_{\text{bare}}$ is phenomenological Stokes friction coefficient and fluid velocity, $\textbf{u}_{\text{fluid}}$ is obtained via interpolation of the flow velocity from the surrounding sites into the monomer position\cite{20}.

LB fluid is characterized by LB parameters. Kinematic viscosity, fluid density, LB grid spacing, LB time step and the coupling constant are fixed as $\nu=3.0$, $\rho=0.864$, $\Delta x_{\text{LB}}=1$, $L_{ij}=\delta_{ij}/\tauLB$, with $\tauLB = 0.05$ and   $\Gamma_{\text{bare}}=20.0$, respectively\cite{18}. The applied electric field is turned on after an equilibration run of duration $t=2\times10^6$ MD steps ($\Delta t=0.01$). In translocation simulations the closed form of the electric field is used \cite{18} but in the second case, the finite element method (FEM) is employed to obtain the electric field by using freeFEM++ package (a partial differential equation solver).

\vspace{1cm}
\section{Results and Discussion}
\vspace{0.5cm}
\subsection{Translocation phenomenon}
In translocation experiments, a voltage bias is used to drive PEs such as DNA through a single nanopore on a substrate. By recording the ionic current modulation due to partially blocking of the channel during polymer translocation, one can  monitor the polymer translocation indirectly\cite{18,24}.

It has been shown previously that the extended electric field in the vicinity of the pore has an important effect on DNA capture and indeed DNA deformation before reaching the pore helps the end monomers of the chain to find the pore entrance more easily\cite{18,25}. Dynamics of charged chain in translocation phenomenon before reaching the pore is a good example to show the importance of polarization effects. 
 
Size of the simulation cell for translocation simulation is $5R_g\times 5R_g\times 10R_g$ , where $R_g$ denotes the gyration radius of the chain. The pore radius and length are $1.5$ and $5$, respectively. The polymer's center of mass is initially located on the pore's axis at a distance equal to the capture radius, $r^*$ from the pore entrance that is defined as the distance at which the polymer is attracted toward the pore due to electrophoretic interactions.

\begin{figure}
\begin{center}
\begin{tabular}{cc}
\raisebox{-15pt}{\includegraphics[width=0.46\columnwidth]{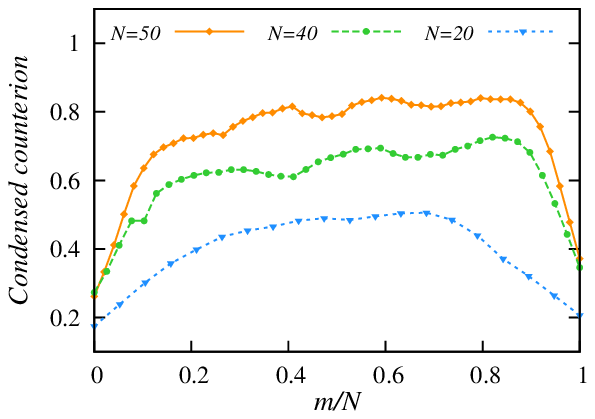}} &
\raisebox{-15pt}{\includegraphics[width=0.48\columnwidth]{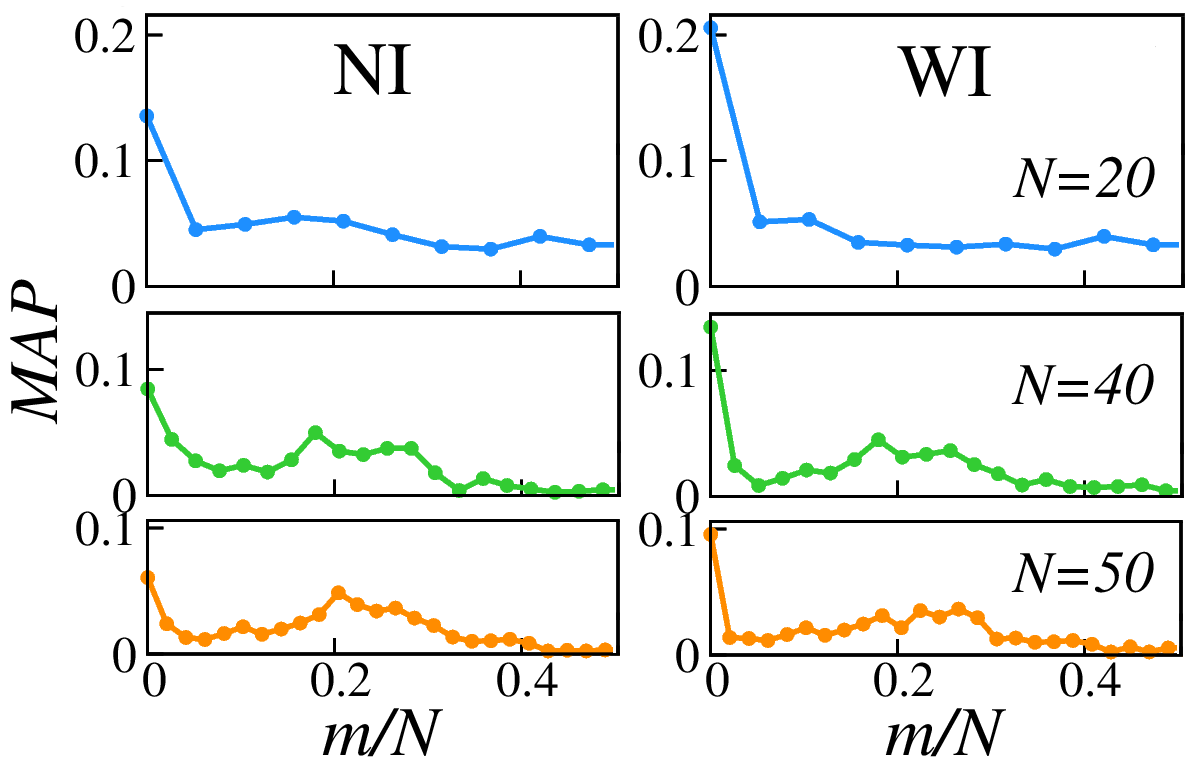}}\\
a & b
\end{tabular}
\end{center}
\vspace*{-15pt}
\caption{\label{counterion}
a) Distribution of condensed CIs along the chain's backbone for different chain lengths, $N$, for the case of $V = 10$ (reduced unit) and no added salt. The monomer ID ($m$) is normalized by the chain length. b) Left panels show the MAP along the chain for different lengths when mean-field effective charges are used instead of explicit CIs (NI). For better comparison between different lengths, we normalized the monomer ID (m) to the degree of polymerization (N). Right panels show the same plots but with explicit CIs (WI).
}
\end{figure}

Distribution of condensed CIs around PEs has been previously studied in equilibrium and in homogeneous electric fields\cite{10,15,18}. Here, we have counted the number of condensed CIs along the chain
in a moderate but inhomogeneous electric field. When the molecule is aligned with the electric field lines, we assign the ID number 0 to the end monomer that is closer to the pore mouth. As it is clear from Fig.~\ref{counterion}-a, the symmetry in the number of condensed CIs along the chain is broken and there are more condensed CIs around the monomers away from the pore.

This increases the polarization effect and leads to a larger effective charge of the closer end monomer, thereby  enhancing the stretching process. The possibility for rearrangement of the CI cloud can, therefore, not be neglected in studies concerned with the dynamics of polyelectrolytes in IEFs.

The importance of this effect is highlighted by introducing a new parameter called MAP($m$) (approaching probability of $m$th monomer), which gives the number of polymer approaches (divided by the total number of approaches) in which the monomers $m$ or $N-m$ are the first monomer of the chain reaching a distance of the pore smaller than 2.5. Figure~\ref{counterion}-b shows MAP versus the normalized monomer ID, $m/N$, for three different chain lengths. The left panels show the results when an effective charge is assigned to the monomers and there is no explicit ion in the system and the right panels show the results of simulations with explicit CIs which can rearrange and polarize the PE complex. As seen from the plots, polarization can enhance the deformation and elongation along the field lines and hence increase the MAP for the end monomers.

\vspace{1cm}
\subsection{Entropic trap}
In the field of electrophoresis in gels and micro and nanofluidic devices, the entropic trapping regime refers to the case where the nominal pore size in the device is commensurate with the radius of gyration of the DNA \cite{1,12} and so PEs in their migration, periodically encounter entropic barriers. So far several designs for microfabricated channels have been introduced and used in experiments. Mobility of PEs and so efficiency and resolution of separation highly depend on the details of the gel, array or channel. Some numerical and theoretical studies discuss that the electric field gradients at the entrance of the slit can suppress the trapping mechanism \cite{26,27}.

\begin{figure}
\begin{center}
\begin{tabular}{cc}
\raisebox{0pt}{\includegraphics[width=.45\columnwidth]{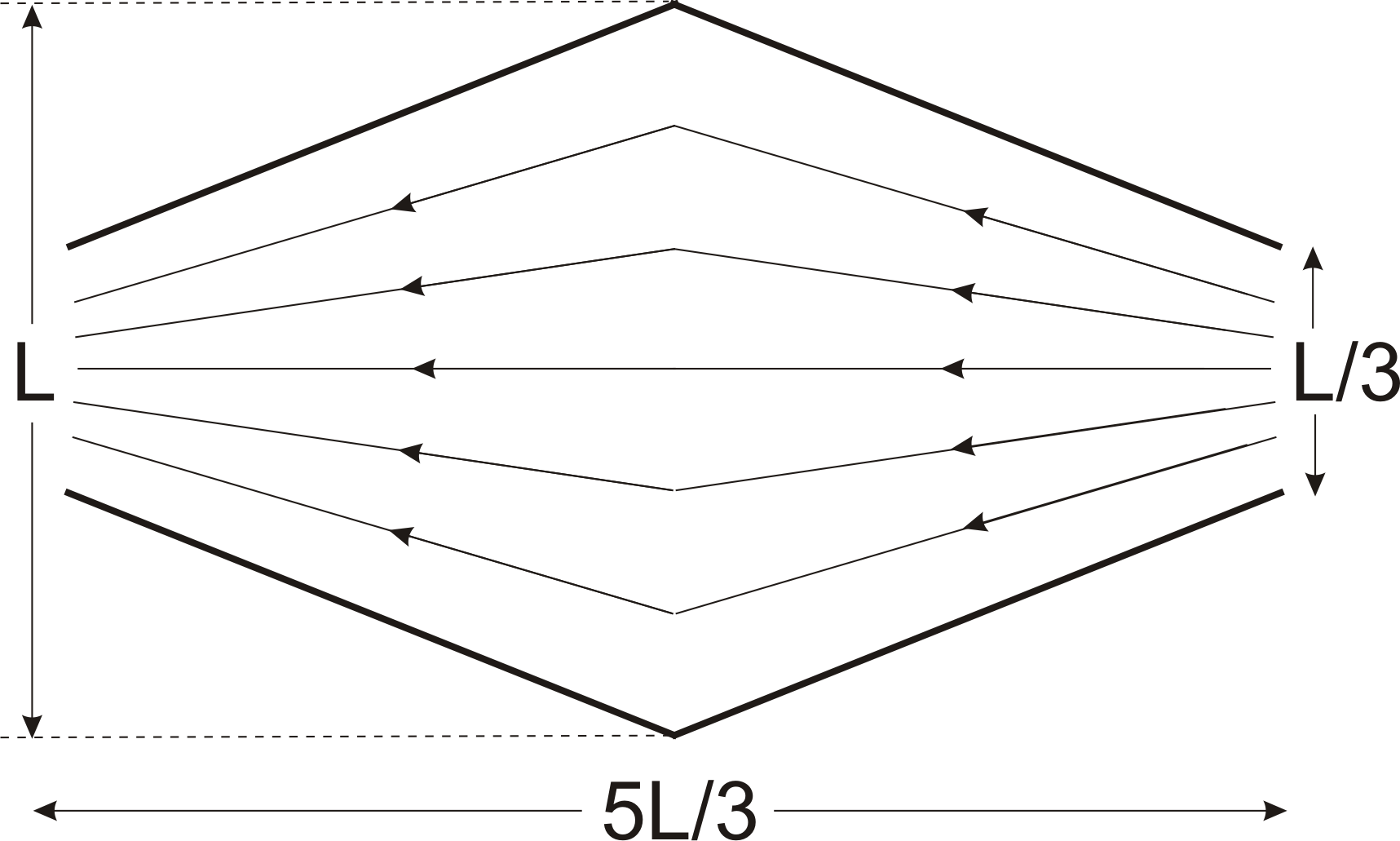}} &
\raisebox{-20pt}{\includegraphics[width=.5\columnwidth]{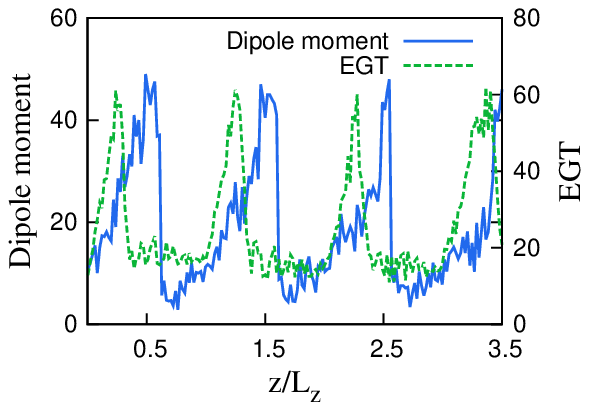}} \\
\raisebox{0pt}{a} & \raisebox{0pt}{b} \\
\raisebox{0pt}{\includegraphics[width=.47\columnwidth]{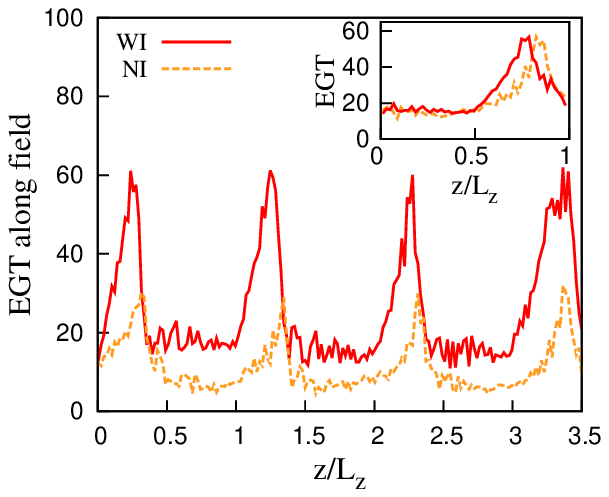}} & \raisebox{0pt}{\includegraphics[width=.47\columnwidth]{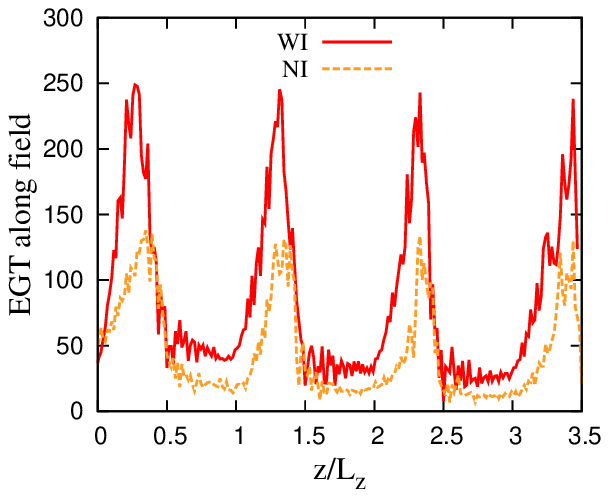}}\\
c & d
\end{tabular}
\end{center} 
\vspace*{-10pt}
\caption{a) Schematic of the channel of simulation of entropic traps. b) Dipole moment of the PE complex versus the position of its center of mass for a polymer of length 30. EGT is plotted as well for comparison. c) EGT of a PE with length 30, projected along the field lines in the position of the center of mass of the chain, versus its position. The inset shows the scaled version of the average of three last cycles d) The same as (c) but for a chain with length 60.\label{field_new}}
\end{figure}

In order to investigate the behavior of PEs in a situation similar to entropic traps, we constructed a periodic simulation box of size $Lx \times Ly \times Lz = L \times L \times 5L/3$ in which wide and narrow regions are obtained by placing conical walls in opposite direction periodically. Figure \ref{field_new}-a shows a schematic view of the box. Such a design may be reminiscent of nano filters that are considered as a subgroup of entropic traps \cite{1}. For each length of the polymer, $L$ is chosen so that the size of the box results in constant monomer concentration of approximately $5mM$.

Simulations are performed for two different lengths and two different situations: 120 simulations without explicit ions (NI) and 120 simulations with CIs and salt (WI). At the beginning, the chain center of mass is located at the middle of the box. After turning on the field, tracing the chain and gathering data continue until the chain's center of mass reaches the end of the box for the fourth time.

Figure~\ref{field_new} depicts for two chain lengths of (c) $N=30$ and (d) $N=60$ the greatest eigenvalue of the gyration tensor (EGT) of the PE (projected along the field lines) versus the $z$ component of the polymer's center of mass position. Apart from different fluctuation amplitudes due to different chain lengths, both plots show that, in the presence of the explicit ions, which promote the polarization of PE complex, chain elongation and alignment are considerably enhanced. The inset of the Fig.~\ref{field_new}-c shows the average of the data over the last three cycles of the motion. Curves are scaled here in order to match the maximum for a better comparison of the temporal sequence. Another important result is that, in the absence of ions, deformation is merely a consequence of inhomogeneity of the field which deforms a charged chain similar an extensional flow. In contrast, a comparison of the delay of deformation in NI and WI cases clearly shows that a PE complex, with its CI cloud, experiences an additional deformation due to polarization.

The effect of CI is further highlighted by plotting in Fig ~\ref{field_new}-b the dipole moment and EGT of a PE complex in the presence of explicit ions (WI). It is seen from this figure that the beginning of the elongation in each cycle nearly coincides with the beginning of the polarization of the chain complex.

\vspace{1cm}
\section{Conclusion}
In summary, we have performed hybrid MD-LB simulations of electrophoretic motion of charged chains in two different setups by taking into account both the electrostatic and hydrodynamic interactions. In the first setup, we investigated the DNA deformation before reaching the pore in translocation phenomenon. By comparing simulations with explicit CIs with those where CIs are replaced by an effective monomer charge, we provided evidence that such a mean-field model for the chain charge significantly underestimates the approaching probability of the end monomers. Non-uniform electric field of a pore is capable of inducing a dipole moment to the PE complex when it is close enough to the pore entrance and this polarization can enhance the deformation and elongation process. 

In the second setup, we investigated the electrophoretic motion of a PE in a conical periodic channel. In two different situations (with and without explicit free ions in the system) deformation and elongation of the chain during its migration in the IEF inside the channel was investigated. We showed that in addition to the extension which is experienced by the chain as a deformable charged molecule in extensional electric field, an additional deformation occurs due to polarization effect of the PE. The contribution of polarization to chain deformation is more sensitive to the gradient of the field. Thus, when a PE is simulated with explicit ions, it would be elongated faster than a bare polyelectrolyte. Charged polyelectrolytes in IEF do, therefore, not exactly behave as a neutral deformable object in extensional flow.

\end{document}